\documentclass[twocolumn,showpacs,preprintnumbers,amsmath,amssymb,prl]{revtex4}

\usepackage{graphicx}

\begin{document}

\title{Exponent Inequalities in Dynamical Systems}
\author{Eytan Katzav $^\dag$ and Moshe Schwartz$^\ddag$}
\affiliation{$^\dag$ Department of Mathematics, King's College London, Strand, London WC2R 2LS, United Kingdom \\
$^\ddag$ Department of Physics, Raymond and Beverly Sackler Faculty of Exact
Sciences, Tel Aviv University, Tel Aviv 69978, Israel}


\begin{abstract}
In this letter we derive exponent inequalities relating the dynamic exponent $z$ to the steady state exponent $\Gamma$ for a general class of stochastically driven dynamical systems. We begin by deriving a general exact inequality, relating the response function and the correlation function, from which the various exponent inequalities emanate. We then distinguish between two classes of dynamical systems and obtain different and complementary inequalities relating $z$ and $\Gamma$. The consequences of those inequalities for a wide set of dynamical problems, including critical dynamics and Kardar-Parisi-Zhang-like problems are discussed.
\end{abstract}


\maketitle
The focus of interest in statistical physics has shifted in the last two decades from equilibrium phase transitions and later the dynamics of phase transitions \cite{HH} to the study of nonequilibrium systems such as various growth models \cite{Mukh97,KPZ,chat99,barabasi95,EW}, front propagations \cite{Golestanian,LeDoussal06,barabasi95}, crack propagation \cite{Katzav06} etc. In spite of that shift the main objects of study remained of a similar nature, a small set of exponents which describe the steady state properties as well as the evolution of the system. Except for the exponents describing critical dynamics and those of a number of one dimensional exactly soluble problems \cite{EW,KPZ,barabasi95,Nonlocal,NMBE,FKPZ,AKPZ}, the sets of exponents given in the literature for many systems belonging to the above categories, vary considerably from author to author and depend strongly on the method of derivation. This is very different from the situation in equilibrium phase transitions, where methods as different as high temperature expansion, momentum space Renormalization-Group (RG) and real space RG yield very close exponents. Under such circumstances rigorous results that can put bounds on the exponents describing the system are obviously most valuable. In the following we present a quite powerful inequality for dynamical stochastic systems, which is an extension of the Schwartz-Soffer inequality derived for quenched random systems \cite{SS}. The inequality is of a generic nature and relates the response at the steady state of some measurable physical field to an external disturbance, to the time dependent correlations of that physical field. This in itself is enough to check approximation schemes or experiments that supply both quantities. In the interesting cases, where the system may be described in terms of a set of exponents, the predictions of the inequality become more dramatic by turning it into an exponent inequality.

Many interesting dynamical physical systems may be described in terms of some physical field, $\phi (\mathbf{r},t)$ driven by a "noise" field, $\eta (\mathbf{r},t)$. The list of systems, described by generic Langevin field equations, includes models of critical dynamics \cite{HH}, growth models of the Kardar-Parisi-Zhang (KPZ) family \cite{KPZ} and its many variants \cite{Mukh97,chat99,barabasi95}, noise driven Navier-Stokes \cite{McComb,ES02} etc.. Strictly speaking, the physical field given as a function of time depends not only on the noise field at earlier times but also on initial conditions. The dependence on initial conditions decays, however, in time, and we are left with an implicit relation between the Fourier transform of the field and the Fourier transform of the noise, $\phi (\mathbf{q},\omega )=\phi \{\mathbf{q},\omega ;\eta (\mathbf{l},\sigma )\}$, where $\eta (\mathbf{l},\sigma )$ is a Gaussian  random field with  $\left\langle {\eta ({\bf{l}},\sigma )} \right\rangle  = 0$, and
\begin{equation}
\left\langle \eta (\mathbf{l},\sigma )\eta (\mathbf{m},\varsigma ) \right\rangle =2{{D}_{0}}(l,\sigma )\delta (\mathbf{l}+\mathbf{m})\delta (\sigma +\varsigma )
 \label{eq:1} \, .
\end{equation}
We are interested in the response function $G(q,\omega)$, to be defined by
\begin{equation}
\left\langle {{\delta \phi ({\bf{q}},\omega )}/{\delta \eta ({\bf{p}},\sigma )}} \right\rangle  \equiv G(q,\omega )\delta
({\bf{q}} - {\bf{p}})\delta (\omega  - \sigma )
 \label{eq:2} \, ,
\end{equation}
and in the correlation function $\Phi(q,\omega)$, defined by
\begin{equation}
\left\langle {\phi ({\bf{q}},\omega )\phi ( - {\bf{p}}, - \sigma )}
\right\rangle  \equiv \Phi (q,\omega )\delta ({\bf{q}} -
{\bf{p}})\delta (\omega  - \sigma )
 \label{eq:3} \, .
\end{equation}
Because of the Gaussian character of the noise, using integration by parts the response function can also be written as
\begin{equation}
G(q,\omega )\delta ({\bf{q}} - {\bf{p}})\delta (\omega  - \sigma ) =
\left\langle {\phi ({\bf{q}},\omega )\eta ( - {\bf{p}}, - \sigma )}
\right\rangle /2D_0 (q,\omega )
 \label{eq:4} \, .
\end{equation}
Note, that if we {\bf define} the response function by the right-hand side of Eq.~(\ref{eq:4}) (which still involves a nontrivial correlation function) the rest of the derivation follows, {\bf even} if the distribution of the noise is {\bf not} Gaussian.

The average $\left\langle \chi (\mathbf{q},\omega )\psi (-\mathbf{p},-\sigma ) \right\rangle$ can be viewed as a scalar product of $\chi (\mathbf{q},\omega )$ and $\psi (\mathbf{p},\sigma )$, since it has all the properties required of a scalar product. Using the Schwartz inequality we find
\begin{eqnarray}
&&\left| G(q,\omega ) \right|\delta (\mathbf{q}-\mathbf{p})\delta (\omega -\sigma ) \\
&&\le \sqrt{\Phi (q,\omega )2{{D}_{0}}(q,\omega )}\delta (\mathbf{q}-\mathbf{p})\delta (\omega -\sigma )/2{{D}_{0}}(q,\omega ) \nonumber
 \label{eq:5} \, .
\end{eqnarray}
Integrating  over $\mathbf{p}$ and $\sigma$ and squaring leads to
\begin{equation}
2|G(q,\omega)|^2 D_0(q,\omega) \le \Phi(q,\omega)
 \label{eq:6} \, .
\end{equation}

The above is a general exact inequality, relating the response function, as defined by the right-hand side of Eq.~(\ref{eq:4}) and the correlation function. To turn that into an exponent inequality, let the equal time correlation be proportional to $q^{-\Gamma}$ for small $q$, i.e. $\Lambda (q) = \int\limits_{ -\infty }^\infty  {d\omega \Phi (q,\omega )} \propto q^{-\Gamma}$, and let the response exponent $\bar z$ characterize the small $q$ behavior of the response function, $|G(q,0)| \propto q^{-\bar z}$. The characteristic frequency, $\omega (q)$, associated with the decay in time of the correlation is given in terms of the dynamic exponent $z$ as \cite{HH} $\omega^{-1}(q)=\pi \Phi (q,0)/\Lambda (q)\propto {{q}^{-z}}$. (Note that in the above we use the traditional extended definition of a power law. A non-negative continuous function, $f(q)$ that vanishes at $q=0$ is said to behave like $q^{\alpha}$ if $\underset{q\to \infty }{\mathop{\lim }}\,\text{ }{{q}^{\beta }}/f(q)=0$ for any $\beta >\alpha$ and $\underset{q\to \infty }{\mathop{\lim }}\,\text{ }f(q)/{{q}^{\gamma }}=0$ for any $\gamma <\alpha $.)

We will concentrate in the following on bare spectral functions $D_0(q,\omega)$ (the noise correlator in Eq.~(\ref{eq:1})) that for small $q$ and $\omega$ have the form
\begin{equation}
 D_0(q,\omega )=B q^{-2\sigma } \qquad \text{for small $q$ and $\omega$}
 \label{eq:7} \, .
\end{equation}
The above form is rich enough to make our point. Still, the discussion of more general cases is straightforward.

The required exponent inequality is obtained now by setting $\omega =0$ in Eq.~(\ref{eq:6}),
\begin{equation}
2\bar z + 2\sigma  \le \Gamma  + z
 \label{eq:8} \, .
\end{equation}
The inequality above relates the response exponent and the dynamic exponent to the static exponent. Note that for a linear system $z=\bar{z}=\Gamma -2\sigma $ and the inequality is exhausted, and satisfied as an equality.

Interestingly, we have observed that most of the important and widely studied dynamical, nonlinear stochastic systems, belong into one of two classes, which will be denoted as class $\text{I}$ and class $\text{II}$ respectively. In each of the classes there exists an additional relation among the exponents, which results in an inequality relating the dynamic exponent $z$ to the static exponent $\Gamma$. The first class to be denoted by $\text{I}$, is that of {\it generalized} Hamiltonian systems. All the classical relaxation models of critical dynamics \cite{HH} belong to class $\text{I}$. A Hamiltonian system is described by a Langevin field equation
\begin{equation}
\gamma \frac{\partial \phi ({\bf{q}})} {\partial t} =  - \frac{\delta H}{\delta \phi(-{\bf{q}})} + \eta ({\bf{q}},t) \quad \text{with} \,\, D(q,\omega ) = D
 \label{eq:9} \, ,
\end{equation}
where the Hamiltonian (free energy functional) is a functional of the physical field $\phi$. It is obvious \cite{HH} that for such systems the following fluctuation-dissipation relation (FDR) holds
\begin{equation}
G(q,0) = \beta \Lambda (q)
 \label{eq:10} \, ,
\end{equation}
where the temperature is given by $k_BT=\beta^{-1}=\frac{D}{\gamma }$.

A generalized Hamiltonian system is described by
\begin{equation}
\gamma \frac{{\partial \phi ({\mathbf{q}})}}{{\partial t}} =  - {\lambda _q}\frac{{\delta H}}{{\delta \phi ( - {\mathbf{q}})}} + \eta ({\mathbf{q}},t)
 \label{eq:11} \, ,
\end{equation}
with  $D(q,\omega )=D \lambda_q$, where $\lambda_q > 0$.
It is easy to prove by setting $\phi_{\mathbf{q}}=\psi_\mathbf{q} \sqrt{\lambda_q}$ and $\eta _\mathbf{q}=\xi_\mathbf{q} \sqrt{\lambda_q}$ and by using relation (\ref{eq:10}) that for the generalized Hamiltonian system the FDR becomes
\begin{equation}
G(q,0) = \beta \lambda_q^{-1} \Lambda _q
 \label{eq:12} \, .
\end{equation}
(If for some $q$'s $\lambda_q$ is zero the corresponding $\phi_{\mathbf{q}}$'s are conserved and may be viewed as parameters in the Hamiltonian rather than dynamical variables.)

Turning the generalized FDR (\ref{eq:12}) into an exponent relation, we obtain for class $\text{I}$
\begin{equation}
\bar z = \Gamma  - 2\sigma
 \label{eq:13} \, .
\end{equation}
Thus, our general exponent inequality (\ref{eq:8}) is turned for class $\text{I}$ into an inequality relating the dynamic and the static exponents,
\begin{equation}
z \ge \Gamma  - 2\sigma  \qquad \text{or equivalently} \quad z \ge \bar z
 \label{eq:14} \, .
\end{equation}

The study of dynamical Hamiltonian systems has a long history and the results are well established. Nevertheless, it is interesting to see how the results derived for the dynamic exponent compare with inequality (\ref{eq:13}). Model A (and C) of Hohenberg and Halperin \cite{HH}, belong to the class discussed above, with $H$ being the  ferromagnetic Landau-Ginzburg-Wilson Hamiltonian (free energy functional) for an $O(n)$ order parameter. They obtain at the transition to second order in $\varepsilon =4-d$, $z=2+c \eta$, where $c$ is a positive constant of order unity and $\eta >0$ is the anomalous dimension. Our inequality yields $z\ge \Gamma =2-\eta $, which is obviously fulfilled by the dynamic exponent given in \cite{HH}. Nevertheless, since $\eta $ is of the order of $\varepsilon^2$, the margin by which the analytic result misses the inequality is rather small. The determination of $c$ for lower dimensions is more difficult \cite{HH,C} yet it seems that the all the derivations of the dynamic exponents yield a non-negative $c$ and therefore obey the inequality.

Model B \cite{HH}, which conserves the order parameter is a generalized Hamiltonian system with a Hamiltonian $H$ identical to that of model A but with $\lambda_q=(q/q_0)^2$. This implies that $\sigma =-1$ and results in $z\ge 4-\eta $. This result should be compared to $z=4-\eta $, which is given in \cite{HH} to all orders in $\varepsilon $. The conclusion here is that in this case the inequality is obeyed as an equality. Interestingly, this is also the case above the upper critical dimension (which is usually $4$) – where $\Gamma =2$, $\bar{z}=2$ and $z=2+2\sigma $ and so the inequality is again saturated.

Class $\text{II}$ is the class of Galilean invariant systems. A myriad of growth models, including the KPZ \cite{KPZ,barabasi95} and MBE families \cite{barabasi95,MBEDRG} as well as the noise driven Navier-Stokes equation \cite{McComb,ES02} and in fact any flow field equation driven by noise that is invariant under Galilean transformations, belong to this class. The stochastic equation governing such systems has the form
\begin{equation}
\gamma \frac{\partial \phi(\mathbf{q})} {\partial t} = F_\mathbf{q} \{\phi\}  + \eta (\mathbf{q},t)
 \label{eq:15} \, ,
\end{equation}
with a generalized force functional,$F_\mathbf{q}$, which for all
$\mathbf{q}$, is independent of $\phi (0)$, i.e. the zeroth Fourier
component. Consequently, it is clear that for members of class
$\text{II}$, $G(0,\omega )={1}/{i\gamma \omega }$ (This may seem
problematic for $\sigma>0$ but can be resolved by considering a
finite system, and taking the limit of an infinite system at the end).
Here, we have to employ the widely used scaling assumption
\cite{Doherty,BC,Moore01} that the response function is given for
small $q$ by
\begin{equation}
G(q,\omega) = \frac{1}{q^{\bar z}} f \left(\frac{\omega}{\omega_q} \right)
 \label{eq:16} \, .
\end{equation}
By definition of $\bar z$, i.e.  $|G(q,0)| \propto q^{-\bar z}$, the scaling function $f(x)$ is a constant for small $x$. On the other hand, because of the form of $G(0,\omega)$, $f(x)$ must obey $f(x)\propto \frac{1}{ix}$ for large $x$. Consequently we obtain $z=\bar{z}$. (A somewhat incomplete derivation of the relation $z=\bar{z}$ appeared in that context \cite{BC} also using the scaling assumption.) The consequence of the relation $z=\bar{z}$ is that the inequality for class $\text{II}$ systems reads
\begin{equation}
z \le \Gamma  - 2\sigma
 \label{eq:17} \, ,
\end{equation}
which is just the opposite inequality to that of class $\text{I}$. (We stress again that the class $\text{II}$ inequality (\ref{eq:17}), stands on less firm grounds than inequality (\ref{eq:14}) for class $\text{I}$ systems, because scaling is assumed in its derivation.)

Many interesting physical systems belong to class $\text{II}$. Among
the more recent are a family of models describing the propagation of
in-plane cracks \cite{Katzav06}, disordered random field elastic
lines \cite{FRG} and a family of models describing the evolution of
wetting lines \cite{Golestanian,LeDoussal06}. In the first two cases
the exponents are given by $z=\Gamma =2$ and
$z=\Gamma-2\sigma=(5-2\sigma)/3$ respectively and in the latter the exponents are
$z=\left( 13+2d \right)/11$ and $\Gamma=(14+3d)/11$, where $d$ is
the dimension of the system. All of these results clearly satisfy the
inequality. Interestingly, in the first two cases the inequality is
saturated and in the last case the results obey the inequality but
the margin is small for $d=1,2,3$.

The most famous systems belonging to class $\text{II}$ are growth
models of the KPZ \cite{KPZ,barabasi95} and MBE families
\cite{MBEDRG,barabasi95}. In those systems, we can obtain more than
just an inequality relating the dynamic exponent $z$ and the steady
state exponent $\Gamma$. This is because the two are related by a
scaling relation $(\Gamma -d)/2+z=2$ for the KPZ family and $(\Gamma
-d)/2+z=4$ for the MBE family \cite{barabasi95}. (Note that $\alpha
=(\Gamma -d)/2$ is the roughness exponent). It is easy to verify
that for KPZ, all the known results for regular KPZ obey the
inequality. Basically, the reason for that is that all methods give
$\Gamma >d$ and $z<2$ even when for $d\ge 2$ the exponents obtained
analytically \cite{KPZ,barabasi95,Doherty,BC,Moore01,Hen91,SE,Canet10} deviate
most considerably from the simulations \cite{KPZsim}. For $d=1$ all
the analytic methods recover the exact results for regular KPZ, so
there is no surprise that for that system the inequality is obeyed. 
This is not the general case, however. As a test
case, we checked if results for the KPZ equation with long-range
interaction \cite{Mukh97,chat99,Nonlocal,NMBE,FKPZ,Hu02,Tang01} obey
the inequality and found \cite{Long} many violations. The only
scheme that did not lead to violation of the inequality is the Self
Consistent Expansion (SCE) \cite{Katzav99,NKPZ03,MBESCE}. It should
be emphasized again that the above statements hold for systems with
long-range nonlinear interactions; however, for systems with long-range
noise, for example, methods like Functional RG \cite{FRG} also
provide results consistent with the inequality. Just to make our
point here, we give, in the following, the results of the one-loop DRG
(Dynamic RG) and show explicitly where they disobey the inequality.

\begin{figure}[ht]
\centerline{\includegraphics[width=6cm]{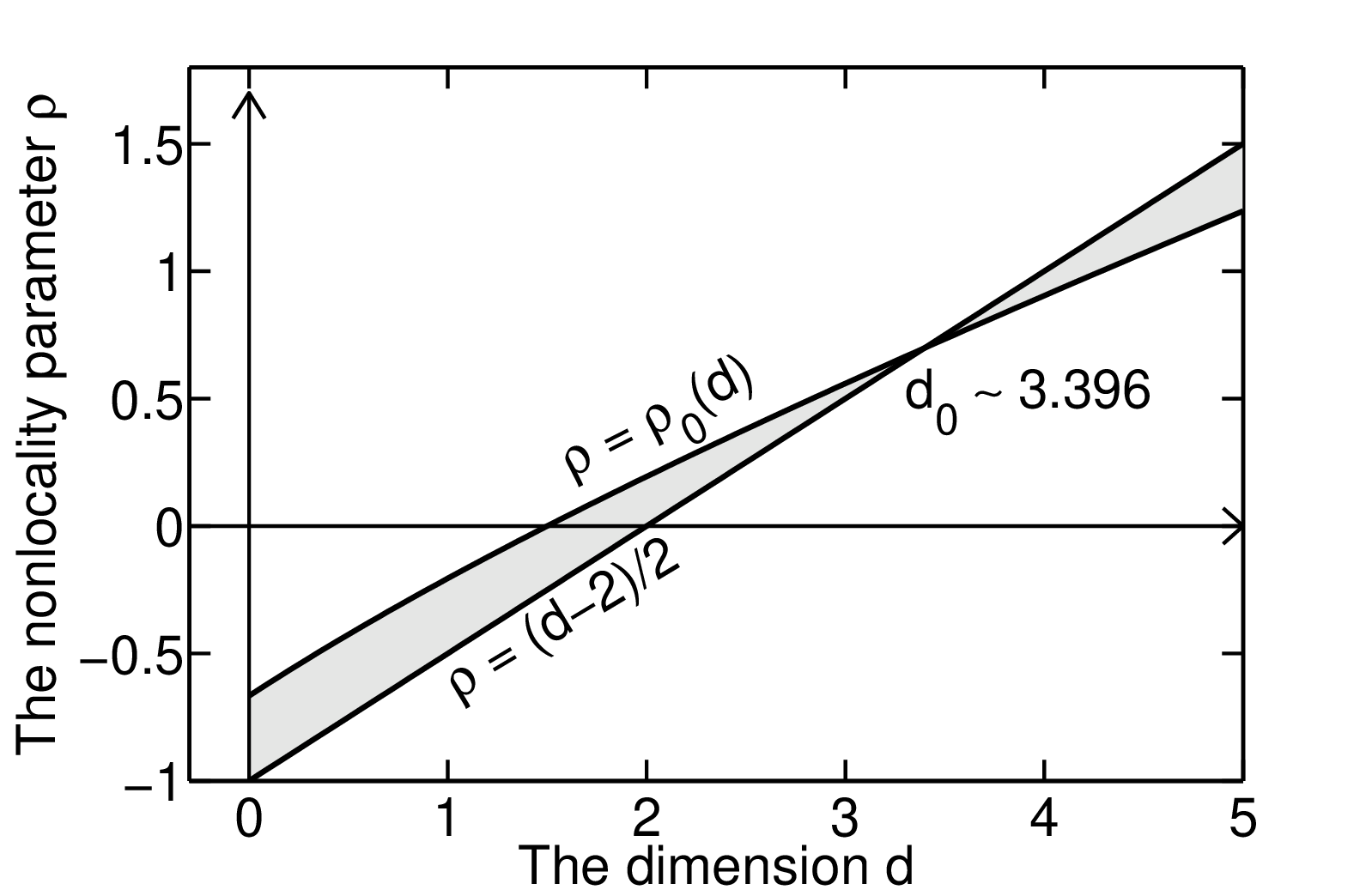}}
 \caption{The violation of the response-correlation inequality (\ref{eq:17}) with $\sigma =0$ by the DRG method derived in Refs.~\cite{Mukh97,chat99} occurs in the shaded area enclosed by $\rho =(d-2)/2$ and $\rho =\rho_0(d)$.}
\label{fig:RG}
\end{figure}

The nonlocal KPZ (NKPZ) equation was introduced in
\cite{Mukh97} to account for the nonlocal hydrodynamic interactions
in the deposition of colloidal particles in a fluid. It was later
generalized to spatially correlated noise ($\sigma \ne 0$) in
Ref.~\cite{chat99}. NKPZ is a generalization of the KPZ system in
which the nonlinear local term of KPZ, $g(\nabla
h(\mathbf{r},t))^2$ is replaced by a nonlocal nonlinear term,
$\frac{1}{2}\int{d \mathbf{r}' g\left( {\mathbf{r}'} \right)\nabla
h\left( \mathbf{r}+\mathbf{r}',t \right)\cdot \nabla h\left(
\mathbf{r}-\mathbf{r}',t \right)}$, where the kernel $g\left(
{\mathbf{r}} \right)$ has a short range part $g_0
\delta^d(\mathbf{r})$ and a long-range part $\sim g_\rho r^{\rho
-d}$ (we call $\rho$ the nonlocality parameter). In order to avoid
an extremely, complicated phase diagram, we discuss here only the
case $g_0=0$. We take also the noise to have $\sigma =0$. The strong
coupling solution found by DRG is \cite{Mukh97}
$z_{DRG}=2+\frac{\left( d-2-2\rho  \right)\left( d-2-3\rho
\right)}{\left( 3+{{2}^{-\rho}} \right)d-6-9\rho }$, where the
scaling relation $(\Gamma-d)/2+z=2-\rho $
\cite{Mukh97,chat99} has been used. The above result violates the
inequality (\ref{eq:17}) over a whole region defined by
$\Gamma _{DRG}-z_{DRG}<0$ marked as shaded in Fig.~\ref{fig:RG}.
In the figure $\rho _0 \left( d \right) = \frac{d -2}{3} + \frac{W\left( \frac{1}{9} 2^{\frac{2-d}{3}} d\ln 2 \right)}{\ln 2}$, where $W(x)$ is the Lambert
function.

For the MBE equation \cite{MBEDRG}, we obtain $z \le (d+8)/3$.
Interestingly, the one-loop DRG result \cite{MBEDRG}, as well as the
Self Consistent Expansion \cite{MBESCE} yield $z = (d+8)/3$, and so
the inequality is saturated.

We could envisage, of course, systems that belong simultaneously to
both classes. In such systems both inequalities (\ref{eq:14}) and
(\ref{eq:17}) combine to give the equality $z = \Gamma  - 2\sigma$.
Reflecting about it, this is the case of model B. More precisely, we
claim that although model B does not belong strictly to class
$\text{II}$, it still effectively belongs to that class. The reason
is that in model B the Hamiltonian $H$ can be replaced by $H^{(0)}$ 
obtained by setting $\phi(0)=0$ in $H$. Using $H^{(0)}$ puts the
dynamical system in class $\text{II}$ as well as in class $\text{I}$. Since the two Hamiltonians $H$ and $H^{(0)}$
are equivalent, at least, in the disordered phase, we conclude that
for model B: $z = 4  - \eta$. Thus the result of Ref.~\cite{HH} for
model B using RG, seems to be exact. Note that the fact that they
obtained this result to all orders in $\varepsilon$ still may be 
consistent with (\ref{eq:14}) holding as a strict inequality. On the
other hand, the proof we present here still depends strongly on a
scaling assumption.

To summarize, in this paper we showed how to generalize the Schwartz-Soffer inequality derived originally for quenched random systems \cite{SS} to dynamical stochastic systems. We show that the inequality, which involves the correlation and the response functions can be translated into a simple inequality relating the scaling exponents $\Gamma$, $z$, $\bar{z}$ and the noise correlation exponent $\sigma$. It turns out that many physical dynamical systems discussed in the literature belong to one of two classes (or to both). In each of these classes there is a (different) additional relation. For class $\text{I}$, i.e. that of Hamiltonian systems, the additional relation is $\bar{z}=\Gamma -2\sigma $. For class $\text{II}$ of Galilean invariant systems, the additional relation is $z=\bar{z}$. These two relations result in two complementary inequalities, $z\ge \Gamma -2\sigma $ for class $\text{I}$ and $z\le \Gamma -2\sigma $ for class $\text{II}$. For systems belonging to both classes we arrive at the conclusion that $z=\Gamma -2\sigma $. We presented a number of cases where the inequality is rather tight or even saturated. We also show that in spite of being extremely simple, the inequality can be quite powerful when examining analytical, numerical and experimental results. To demonstrate the utility of the inequality from that aspect, we reviewed analytical results for the nonlocal KPZ model, by one method: the Dynamical Renormalization Group which is shown to disobey the inequality for a whole range of parameters. This has an important implication on the choice of analytical tools when dealing with such stochastic models.

It would be surprising if the two classes above cover all the possible dynamical systems. An open question remains if there exist other classes, involving stochastic driving, which are relevant in describing physical systems. Another open question is whether there are more systems of interest that belong to classes $\text{I}$ and $\text{II}$ simultaneously.
Hopefully, the simplicity yet strength of this result will motivate researchers to explore the usefulness of rigorous inequalities and derive improved ones alongside the more popular chase of approximate equalities.

\end{document}